\UseRawInputEncoding

\documentclass[runningheads]{llncs}

\usepackage[T1]{fontenc}
\usepackage{graphicx}
\usepackage{booktabs}
\usepackage{amsmath,amssymb}
\usepackage{xspace}
\usepackage{listings}
\usepackage{microtype}
\usepackage{url}
\usepackage{hyperref}
\usepackage{float}

\usepackage{booktabs}

\usepackage{amssymb}

\usepackage{xcolor}
\usepackage{listings}

\usepackage{comment}

\definecolor{LeanKeyword}{RGB}{28,58,255}
\definecolor{LeanComment}{RGB}{0,120,70}
\definecolor{LeanBackground}{RGB}{248,248,248}
\definecolor{LeanFrame}{RGB}{180,180,180}

\lstdefinelanguage{Lean}{
  morekeywords={
    def,theorem,lemma,example,structure,inductive,where,
    deriving,namespace,end,if,then,else,match,with,
    let,in,by,exact,unfold,cases,rfl,rw,simp,apply,
    intro,fun,Type,Prop
  },
  sensitive=true,
  morecomment=[l]{--},
  morecomment=[s]{/-}{-/},
  morestring=[b]"
}

\newcommand{\framework}{\textsc{LeanSafeHRC}\xspace}
\newcommand{\lean}{Lean~4\xspace}

\title{\framework: Verified Executable Runtime Safety Monitoring
for Human-Robot Collaboration}

\titlerunning{Verified Runtime Safety Monitoring for HRC}

\author{Adnan Rashid\orcidID{0000-0002-9729-584X}}

\authorrunning{A. Rashid}

\institute{
School of Electrical Engineering and Computer Science (SEECS)\\
National University of Sciences and Technology (NUST)\\
Islamabad, Pakistan\\
\email{adnan.rashid@seecs.nust.edu.pk}
}

\begin{document}

\maketitle


\begin{abstract}

Runtime safety monitors make safety-critical decisions during human-robot collaboration, yet their decision logic is typically implemented as conventional software and validated primarily through testing and simulation.
We present \framework, a modular \lean framework that formalizes runtime monitoring as an executable pipeline from runtime observations and safety configurations to qualitative risk assessment, verified policy evaluation, canonical action selection, and monitor execution.
Machine-checked proofs establish policy soundness, deterministic and risk-responsive behaviour, prevention of unsafe continuation, and minimal intervention. A representative collaborative assembly case study demonstrates direct execution of the verified monitor across proximity, velocity, force, and emergency scenarios.
The results show that executable runtime monitoring and interactive theorem proving can be integrated within a single Lean~4 development, providing rigorous correctness guarantees for safety-critical human-robot collaboration.


\end{abstract}


\section{Introduction} \label{sec:introduction}

Human-Robot Collaboration (HRC)~\cite{singh2026human} is increasingly
adopted in manufacturing, healthcare, logistics, and service robotics,
where humans and autonomous robots operate within shared workspaces.
Ensuring safe interaction requires runtime monitors that continuously
assess operating conditions and initiate appropriate mitigation actions
whenever unsafe situations arise. Since these monitors directly influence
safety-critical decisions, their correctness is of fundamental importance.

Conventional assurance of robotic and runtime-monitoring systems commonly
relies on testing, simulation, and experimental evaluation
~\cite{araujo2023testing,desai2019soter,kozamernik2023visual}. These
techniques provide valuable empirical evidence by evaluating system
behaviour under representative operating conditions. However, they can
exercise only a subset of the possible behaviours captured by the
underlying model and therefore cannot, by themselves, establish correctness
for all admissible executions.

Formal methods provide a complementary means of establishing stronger
correctness guarantees. Model-checking techniques have been applied to
verify interaction protocols, controller behaviour, and temporal safety
properties of collaborative robotic systems
~\cite{AraizaIllan2015,Webster2020}. Interactive theorem proving further
supports machine-checked reasoning over richer mathematical and robotic
models without relying on exhaustive state exploration. Existing
theorem-proving developments have formalized and verified robotic
kinematics, robot software and system models, and physical human-robot
interaction~\cite{wang2023formalization,yan2026automated,Abed2026}.
Nevertheless, these approaches largely focus on the offline verification
of robotic models and algorithms rather than on constructing executable
runtime monitors whose decision logic is itself accompanied by
machine-checked correctness proofs.

Against this background, recent research has developed a range of runtime
monitoring and safety-supervision techniques for autonomous and robotic
systems~\cite{Yel2020,Rahman2021,Lan2024,RTAMT2024}. These approaches
provide practical mechanisms for detecting hazardous conditions during
system execution and initiating appropriate safety responses. However,
despite their practical effectiveness, the monitoring logic is generally
not developed as an executable artifact together with machine-checked
correctness proofs within an interactive theorem prover. Consequently, a
gap remains between practical runtime monitoring and deductively verified
monitor implementation.

Table~\ref{tab:comparison} summarizes this distinction across representative formal verification and runtime monitoring approaches. Formal verification methods generally provide strong correctness guarantees for offline robotic models, whereas runtime-monitoring approaches emphasize executable safety supervision during operation. The comparison highlights the opportunity to combine executable monitoring and interactive theorem proving within a single formal development.

\vspace*{-0.5cm}

\begin{table}[H]
\centering
\caption{Comparison of Representative Formal Verification and Runtime-Monitoring Approaches for Robotic Systems}
\label{tab:comparison}
\renewcommand{\arraystretch}{0.9}
\setlength{\tabcolsep}{4pt}
\small
\begin{tabular}{p{3.2cm}p{1.2cm}p{1.2cm}p{1.2cm}p{1.2cm}p{1.2cm}p{1.0cm}}
\toprule
\textbf{Approach}
&
\textbf{HRC}
&
\textbf{RM}
&
\textbf{FV}
&
\textbf{ITP}
&
\textbf{Exec.}
&
\textbf{Unified}
\\
\midrule

Webster et al.~\cite{Webster2016}
&
Yes & No & Yes & No & No & No
\\

Webster et al.~\cite{Webster2020}
&
Yes & No & Yes & No & No & No
\\

Rahman et al.~\cite{Rahman2021}
&
Partial & Yes & Partial & No & Yes & No
\\

Yel et al.~\cite{Yel2020}
&
Partial & Yes & Partial & No & Yes & No
\\

Abed et al.~\cite{Abed2026}
&
Yes & No & Yes & Yes & Partial & No
\\

\textbf{\framework}
&
Yes & Yes & Yes & Yes & Yes & Yes
\\

\bottomrule
\end{tabular}

\vspace{1mm}
\footnotesize
RM = runtime monitoring;
FV = formal verification;
ITP = interactive theorem proving;
Exec. = executable monitor;
Unified = executable implementation and machine-checked proofs developed
within one formal environment;
Partial = indicates that the corresponding aspect is addressed indirectly
or only for a restricted component of the overall approach.
\end{table}

\vspace*{-0.6cm}

To address this gap, we present~\framework, a Lean~4 framework for verified runtime safety monitoring in human-robot collaboration. The framework models the complete monitoring pipeline, including priority-based runtime risk assessment, policy evaluation, canonical action selection, and monitor construction, as executable Lean definitions. These definitions simultaneously serve as the executable monitoring
algorithm and the basis for machine-checked correctness proofs, thereby avoiding a separate monitor implementation whose correspondence with the verified model would require additional justification.

Lean~4 is particularly suitable for this work because it combines executable functional programming and machine-checked theorem proving within a single environment. Consequently, the same definitions can serve both as the runtime-monitoring implementation and as the objects of formal verification, avoiding a separate implementation layer whose correspondence with the verified model would require additional justification. Its support for inductive datatypes, executable definitions, and machine-checked reasoning also naturally accommodates the discrete risk levels, safety policies, and decision procedures used in \framework.

The framework is evaluated through a representative collaborative assembly case study covering normal operation, proximity-based hazards, operational safety violations, and emergency conditions. The observed monitor behaviour agrees with the machine-checked correctness guarantees established within the formal development.

The main contributions of this paper are as follows.

\vspace*{-0.2cm}

\begin{itemize}
\item[$\circ$] To the best of our knowledge, we present the first Lean~4 framework that integrates executable runtime monitoring with machine-checked verification for human-robot collaboration (Section~\ref{sec:framework}).

\item[$\circ$] We formalize a modular monitoring architecture comprising priority-based runtime risk assessment, verified policy evaluation, canonical action selection, and executable monitor construction (Section~\ref{sec:monitoring}).

\item[$\circ$] We establish machine-checked correctness guarantees covering policy soundness, deterministic and risk-responsive behaviour, prevention of unsafe continuation, and minimal intervention (Section~\ref{sec:correctness}).

\item[$\circ$] We demonstrate the framework through an executable
collaborative assembly case study showing that runtime execution is
consistent with the machine-checked correctness properties (Section~\ref{sec:case-study}).
\end{itemize}




\vspace*{-0.4cm}

\section{\framework Framework}
\label{sec:framework}

\vspace*{-0.2cm}

The objective of \framework\ is to represent runtime safety monitoring as a unified executable and machine-checked formal development. Runtime observations, risk assessment, policy evaluation, action selection, and
monitoring decisions are represented directly in Lean~4, providing the formal basis for both execution and the correctness properties established later in the paper.


\vspace*{-0.2cm}

\subsection{Design Philosophy}
\label{sec:framework-design}

The framework is guided by three complementary design principles.

\textbf{Executable formalization.}
The monitoring logic is defined as executable Lean functions rather than abstract specifications. These functions can be evaluated directly on instances of the runtime state while simultaneously serving as the subject of machine-checked correctness proofs. Consequently, implementation and verification remain synchronized throughout the development.

\textbf{Layered decomposition.}
The monitoring process is organized into independent layers comprising
runtime observations, risk assessment, policy evaluation, action
selection, and monitor execution. The explicit separation of these
components improves modularity and enables localized reasoning about
individual stages of the monitoring pipeline.

\textbf{Platform independence.}
The framework captures only the information required for runtime safety
decision making. Application-specific measurements are supplied through
a generic runtime-state interface, while safety thresholds are provided
through a separate configuration. This allows the same formalization to
be instantiated across different collaborative robotic applications
without modifying the underlying monitoring architecture.


\vspace*{-0.3cm}

\subsection{Layered Architecture}
\label{sec:framework-architecture}

\begin{figure}[H]
\centering
\includegraphics[width=0.80\textwidth]
  {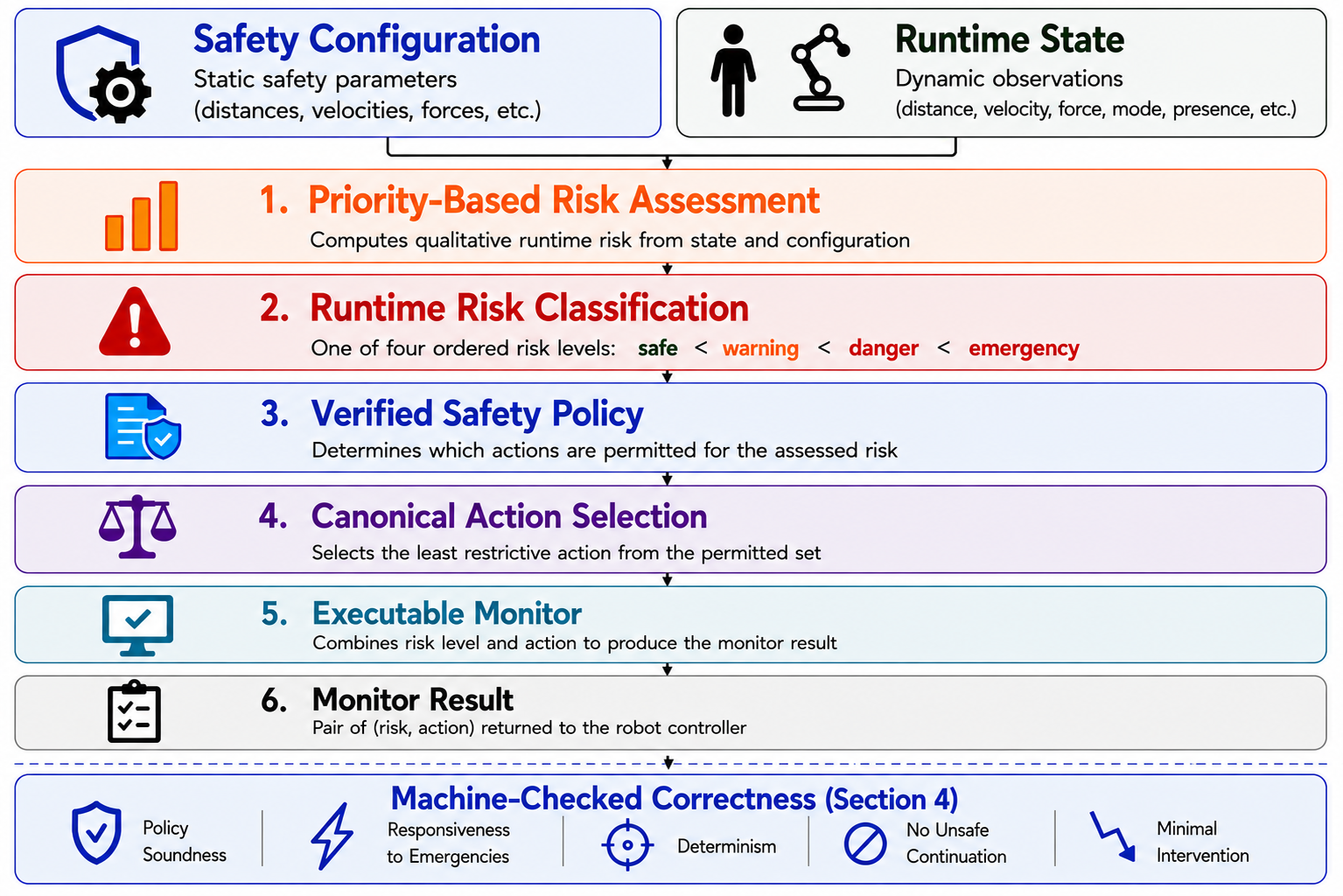}
\vspace*{-0.2cm}
\caption{Layered Architecture of \framework}
\label{fig:framework-architecture}
\end{figure}

Figure~\ref{fig:framework-architecture} illustrates the layered architecture of \framework. The framework transforms a static safety configuration and the current HRC runtime state into a verified monitoring decision through four successive stages, namely risk assessment, policy evaluation, canonical action selection, and executable monitor construction. Each stage is represented by executable Lean definitions, allowing the complete monitoring pipeline to be executed and formally
reasoned about within a single development.

\vspace*{-0.3cm}

\vspace*{-0.3cm}

The monitoring process begins with two complementary inputs. The \textit{safety configuration} stores the static distance, velocity, and force thresholds that define the operating constraints of the collaborative system, whereas the \textit{runtime state} captures the
current observations obtained from the robot and its environment. Together, these inputs provide the information required for runtime decision making.

Risk assessment maps the current runtime state to one of four ordered qualitative risk levels: \texttt{safe}, \texttt{warning}, \texttt{danger}, or \texttt{emergency}. The assessed risk is then
interpreted by the verified safety policy, which determines the actions permitted under the current operating conditions. A canonical action selection procedure subsequently chooses the least restrictive
policy-compliant intervention, and the executable monitor combines the computed risk level and selected action into the final monitoring result.

The lower part of Figure~\ref{fig:framework-architecture} summarizes the principal correctness properties established later in the paper. These guarantees include policy soundness, responsiveness,
determinism, prevention of unsafe continuation, and minimal intervention. Collectively, they establish that the executable monitor not only produces monitoring decisions but also satisfies the formally verified safety properties proved in Section~\ref{sec:correctness}.


\vspace*{-0.4cm}

\subsection{Framework Components}
\label{sec:framework-components}

\vspace*{-0.2cm}

The framework separates static safety parameters from dynamic runtime observations. This distinction allows a single safety configuration to be reused across multiple monitoring cycles while the runtime state is updated continuously during execution.
The static configuration is represented by the parameterized structure
\(\mathit{SafetyConfig}\;\alpha\) as follows:

\vspace*{-0.2cm}

\[
\mathit{SafetyConfig}\;\alpha =
\left\{
\begin{array}{ll}
\mathit{warningDistance} &: \alpha,\\
\mathit{stopDistance} &: \alpha,\\
\mathit{maxVelocity} &: \alpha,\\
\mathit{maxForce} &: \alpha.
\end{array}
\right.
\]

These parameters define the distance, velocity, and interaction-force limits used throughout runtime risk assessment. The numerical type $\alpha$ is kept generic rather than fixed to $\mathbb{R}$, since the monitoring logic relies primarily on ordered comparisons with safety
thresholds; this permits the same formalization to be instantiated with different numerical representations without changing the monitoring architecture. 
For a physically meaningful distance configuration, the thresholds are intended to satisfy $ \mathit{stopDistance} < \mathit{warningDistance}$ so that the protective-stop region lies strictly inside the warning region. The current operating conditions are captured by the runtime state

\vspace*{-0.1cm}

\[
\mathit{HRCState}\;\alpha =
\left\{
\begin{array}{ll}
\mathit{distance} &: \alpha,\\
\mathit{velocity} &: \alpha,\\
\mathit{interactionForce} &: \alpha,\\
\mathit{humanPresence} &: \mathit{HumanPresence},\\
\mathit{emergencySignal} &: \mathit{Bool},\\
\mathit{mode} &: \mathit{RobotMode}.
\end{array}
\right.
\]

Here, $\mathit{distance}$ denotes the current human--robot separation distance, $\mathit{velocity}$ denotes the current robot velocity, and $\mathit{interactionForce}$ denotes the measured human--robot interaction force. The remaining fields record human presence, emergency status, and the current robot operating mode.

The principal formal entities employed throughout the framework are summarized in Table~\ref{tab:framework-types}.

\vspace*{-0.3cm}

\begin{table}[H]
\centering
\caption{Principal Formal Entities in \framework}
\label{tab:framework-types}
\footnotesize
\renewcommand{\arraystretch}{0.9}
\setlength{\tabcolsep}{4pt}
\begin{tabular}{p{2.8cm}p{8.4cm}}
\toprule
\textbf{Formal Entity} & \textbf{Purpose} \\
\midrule

\texttt{RobotMode}
&
Represents the robot operating mode
(\texttt{idle}, \texttt{automatic},
\texttt{collaborative},
\texttt{protectiveStop},
\texttt{emergencyStop}).
\\

\texttt{HumanPresence}
&
Indicates whether a human is present in the monitored workspace.
\\

\texttt{RiskLevel}
&
Represents the ordered qualitative risk levels
(\texttt{safe},
\texttt{warning},
\texttt{danger},
\texttt{emergency}).
\\

\texttt{MonitorAction}
&
Represents the monitor interventions
(\texttt{continueOperation},
\texttt{reduceSpeed},
\texttt{protectiveStop},
\texttt{emergencyStop}).
\\

\texttt{SafetyConfig} \(\alpha\)
&
Stores the static safety thresholds.
\\

\texttt{HRCState} \(\alpha\)
&
Stores the current runtime observations.
\\

\bottomrule
\end{tabular}
\end{table}

\vspace*{-0.4cm}

Consequently, every execution of the monitoring framework is determined by a configuration-state pair

\[
(\mathit{SafetyConfig}\;\alpha,\;
 \mathit{HRCState}\;\alpha)
\longrightarrow
\mathit{MonitorResult},
\]

\noindent where the intermediate risk classification and selected monitor action
remain explicit, allowing both stages to be executed independently and referenced directly in the correctness proofs presented in Section~\ref{sec:correctness}.

For conciseness, the paper presents only the principal definitions required to explain the monitoring architecture. The complete Lean~4 development, including the underlying datatypes, auxiliary definitions, executable functions, and machine-checked proofs, is available in the accompanying source-code repository\footnote{\url{https://github.com/codeforpapers3/leansafehrc}}.


\vspace*{-0.2cm}

\subsection{Ordered Risk and Action Domains}
\label{sec:framework-orderings}

The framework assigns explicit severity measures to both runtime risk levels and monitor actions. These orderings provide the mathematical foundation for risk escalation during runtime assessment and for the minimal-intervention property established later in the paper.

The framework assigns
$\texttt{safe}\mapsto0$,
$\texttt{warning}\mapsto1$,
$\texttt{danger}\mapsto2$, and
$\texttt{emergency}$ $\mapsto3$,
yielding the ordering
$\texttt{safe}<\texttt{warning}<\texttt{danger}<\texttt{emergency}$.

This ordering enables the risk-assessment procedure to compare and prioritize multiple safety violations, with higher-severity violations taking precedence over lower-severity ones.

Monitor actions are ordered independently according to intervention
severity:

\vspace*{-0.2cm}
\[
\begin{aligned}
\mathit{actionSeverity}(\texttt{continueOperation}) &= 0,\\
\mathit{actionSeverity}(\texttt{reduceSpeed})       &= 1,\\
\mathit{actionSeverity}(\texttt{protectiveStop})    &= 2,\\
\mathit{actionSeverity}(\texttt{emergencyStop})     &= 3.
\end{aligned}
\]

\vspace*{-0.2cm}

This induces the ordering
\texttt{continueOperation} $<$ \texttt{reduceSpeed} $<$
\texttt{protectiveStop} $<$ \texttt{emergencyStop}.
Unlike the risk ordering, this action ordering represents increasing intervention severity and provides the basis for selecting the least restrictive policy-compliant action and establishing the minimal-intervention theorem in Section~\ref{sec:correctness}.

\section{Executable Runtime Monitoring}
\label{sec:monitoring}

\vspace*{-0.2cm}

Building upon the framework introduced in Section~\ref{sec:framework}, \framework\ realizes runtime safety monitoring as a sequence of executable
Lean~4 definitions. This section develops the principal components of this pipeline, from runtime risk assessment and policy evaluation to canonical action selection and construction of the final monitoring result.



\subsection{Monitoring Workflow}
\label{sec:workflow}

\begin{figure}[t]
\includegraphics[width=1.00\textwidth]{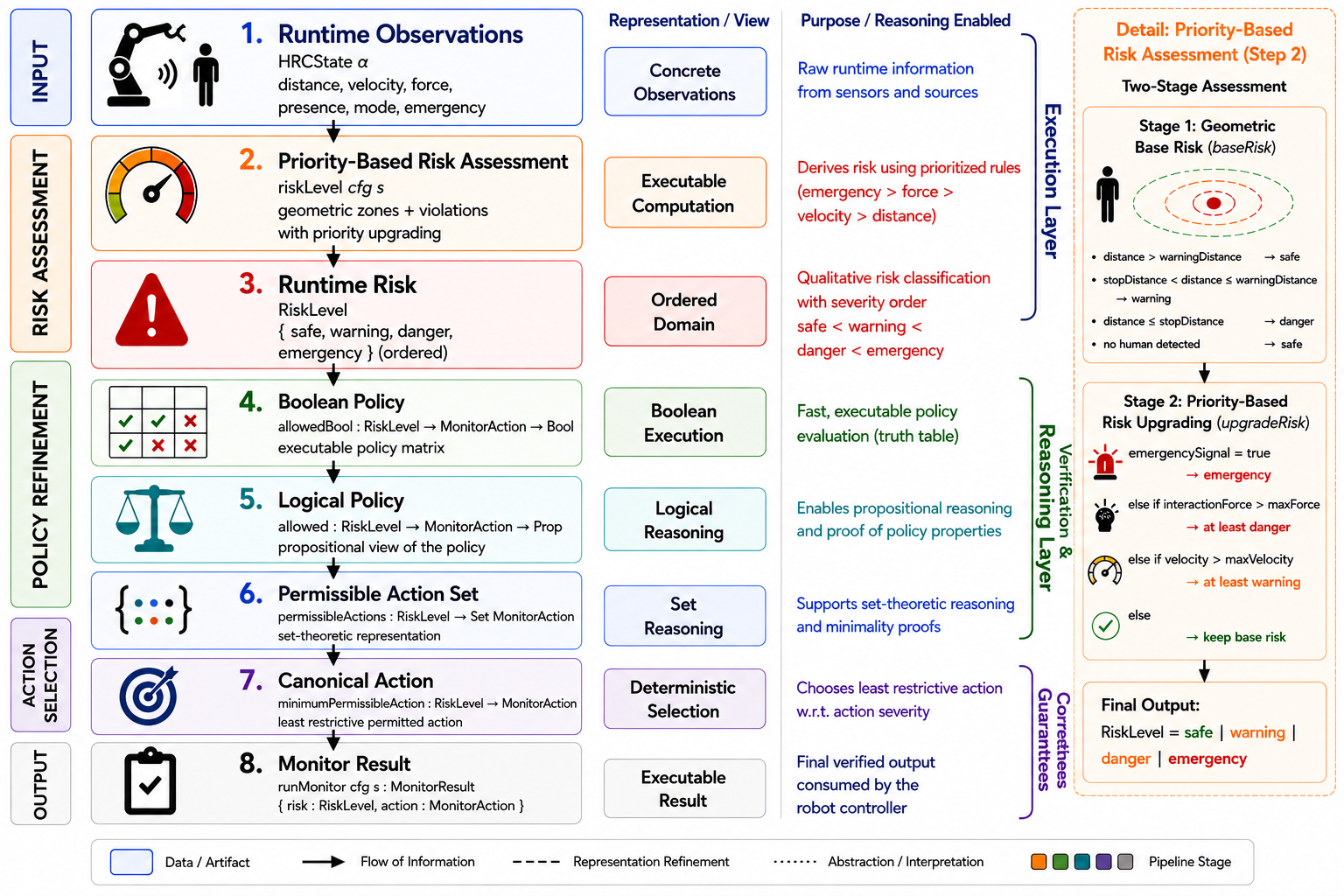}
\caption{Executable monitoring workflow in \framework. 
}
\label{fig:monitor-workflow}
\end{figure}

Figure~\ref{fig:monitor-workflow} presents the executable monitoring
workflow of \framework. Rather than relying on a single
monolithic decision procedure, the framework decomposes runtime
monitoring into four executable stages: runtime risk assessment, policy
evaluation, canonical action selection, and monitor construction. Each
stage consumes the output of its predecessor while exposing a well-defined
interface for both execution and formal reasoning.

Runtime monitoring begins with the current HRC state together with the
corresponding safety configuration. The risk-assessment procedure first
computes one of four ordered qualitative risk levels:
\texttt{safe}, \texttt{warning}, \texttt{danger}, or
\texttt{emergency}. The verified safety policy then determines the
actions permitted for the assessed risk, after which the canonical
action-selection procedure chooses the least restrictive
policy-compliant intervention. Finally, the executable monitor combines
the computed risk level and selected action into the monitoring result
that can be supplied to the robot controller.
In the present framework, runtime monitoring refers to the executable evaluation of the monitor over observations supplied through \texttt{HRCState}; acquisition of these observations from physical sensors or robotic middleware lies outside the scope of the current formalization.

The workflow deliberately decomposes runtime decision making into separate executable stages while retaining implementation and verification within the same formal development. Risk assessment, policy evaluation, action selection, and monitor construction can therefore be evaluated and reasoned about individually while remaining part of a single verified monitoring pipeline.
The following subsections describe these stages in detail, beginning with priority-based runtime risk assessment.



\subsection{Priority-Based Risk Assessment}
\label{sec:risk-assessment}

The first stage of the monitoring pipeline computes the qualitative
runtime risk associated with the current HRC state. The assessment
combines geometric proximity with operational safety constraints using a
two-stage procedure consisting of base-risk computation followed by
priority-based risk escalation.

The geometric assessment classifies the workspace according to the relative separation between the human and the robot. Depending on the measured distance and the configured safety thresholds, the resulting base risk is classified as \texttt{safe}, \texttt{warning}, or
\texttt{danger}. This initial classification is subsequently refined by considering additional safety indicators. Emergency conditions immediately produce the highest risk level, whereas force and velocity violations raise the current classification only when a higher level of risk is required. Consequently, the final assessment preserves the geometric interpretation
while incorporating operational hazards according to their prescribed priorities.

The complete executable classifier is defined as

\[
\operatorname{riskLevel}(c,s)
=
\operatorname{upgradeRisk}
\bigl(
c,
s,
\operatorname{baseRisk}(c,s)
\bigr),
\]

where \(c\) denotes the safety configuration and \(s\) the current
runtime state.

Listing~\ref{lst:riskLevel} presents the executable implementation. The monitor first computes the geometric base risk and then applies the priority-based upgrading procedure to obtain the final risk classification.


\begin{lstlisting}[caption={Executable priority-based runtime risk assessment.},
label={lst:riskLevel}]
def riskLevel (cfg : SafetyConfig α)
              (s : HRCState α) : RiskLevel :=
  upgradeRisk cfg s (baseRisk cfg s)
\end{lstlisting}

The Lean development further establishes that the upgrading procedure is monotonic with respect to risk severity: additional safety violations may preserve or increase the current classification but cannot reduce it. This property supports the correctness results established in Section~\ref{sec:correctness}.



\subsection{Runtime Safety Policy}
\label{sec:policy}

The qualitative risk level produced by the assessment procedure is
interpreted through a verified runtime safety policy. Rather than
associating each risk level with a single action, the policy specifies
the complete set of actions permitted for each operating condition.
Action selection is therefore separated from policy specification,
allowing admissible actions to be determined independently of the
mechanism used to select a particular intervention.

The executable policy is summarized in
Table~\ref{tab:policy}. As the assessed risk increases, the policy
becomes progressively more restrictive by eliminating actions that no
longer satisfy the required level of safety. Consequently, emergency
conditions permit only the emergency-stop action, whereas warning and
danger conditions retain intermediate mitigation strategies appropriate
to their respective severity levels.

\vspace*{-0.4cm}

\begin{table}[H]
\centering
\caption{Executable Runtime Safety Policy (\texttt{allowedBool}).}
\label{tab:policy}
\footnotesize
\renewcommand{\arraystretch}{0.9}
\setlength{\tabcolsep}{4pt}
\begin{tabular}{p{2.7cm}p{2.2cm}p{1.5cm}p{2.0cm}p{2.0cm}}
\toprule
\textbf{Runtime \ \ \ \ \ \ \ \ \ Risk}
&
\textbf{Continue \ \ \ \ \ \ \ Operation}
&
\textbf{Reduce Speed}
&
\textbf{Protective Stop}
&
\textbf{Emergency Stop}
\\
\midrule

Safe
&
\checkmark
&
\checkmark
&
\checkmark
&
\checkmark
\\

Warning
&
--
&
\checkmark
&
\checkmark
&
\checkmark
\\

Danger
&
--
&
--
&
\checkmark
&
\checkmark
\\

Emergency
&
--
&
--
&
--
&
\checkmark
\\

\bottomrule
\end{tabular}
\end{table}

\vspace*{-0.4cm}

Here, permissibility denotes safety admissibility rather than operational
preference. Thus, an action that is more restrictive than necessary may
remain permissible at a lower risk level; for example, a protective stop
or emergency stop is not prohibited when the assessed risk is
\texttt{safe}. The canonical action-selection procedure subsequently
resolves this choice by selecting the least restrictive permissible
action, as described in Section~\ref{sec:monitor-construction}.

For efficient execution, the policy is implemented as the Boolean function \texttt{allowedBool}. For formal reasoning, the framework derives two equivalent views: the logical relation $\texttt{allowed}(r,a)\equiv\texttt{allowedBool}(r,a)=\texttt{true}$,
and the set of admissible actions
$\texttt{permissibleActions}(r)=\{\,a\mid\texttt{allowed}(r,a)\,\}$, where \(r\) denotes the assessed runtime risk and \(a\) a candidate
monitor action.
Machine-checked equivalences connect Boolean evaluation, logical reasoning, and set membership. Thus, monitor execution uses the Boolean policy while the correctness proofs can reason propositionally or set-theoretically without introducing an independent policy specification.
The verification therefore establishes correctness and compliance with the formally specified policy; application-specific validation of the selected safety thresholds and mitigation requirements remains the responsibility of the particular HRC system instantiation.


\vspace*{-0.3cm}

\subsection{Canonical Action Selection and Monitor Construction}
\label{sec:monitor-construction}

The executable monitor is obtained by composing the risk-assessment and policy layers developed in the preceding subsections. At the monitor level, \texttt{monitorRisk} provides a direct interface to the risk classifier,
with $\texttt{monitorRisk}(c,s)=\texttt{riskLevel}(c,s)$. Given the assessed runtime risk, the monitor selects a unique canonical action and combines
both results into the final monitoring decision.

Rather than searching the set of permissible actions during execution,
the framework associates each runtime risk with the least restrictive
policy-compliant action:

\vspace*{-0.3cm}

\[
\begin{aligned}
\texttt{safe}
      &\longmapsto
      \texttt{continueOperation},\\
\texttt{warning}
      &\longmapsto
      \texttt{reduceSpeed},\\
\texttt{danger}
      &\longmapsto
      \texttt{protectiveStop},\\
\texttt{emergency}
      &\longmapsto
      \texttt{emergencyStop}.
\end{aligned}
\]

This deterministic correspondence provides an executable canonical choice for each risk level and forms the basis for the minimal-intervention property proved in Section~\ref{sec:correctness}.

Listing~\ref{lst:monitor} shows the executable Lean implementation. The canonical action selected for each qualitative risk level is composed with the assessed risk to construct the complete runtime monitor.

\begin{lstlisting}[
caption={Executable composition of the runtime monitor.},
label={lst:monitor}]
def monitorAction (cfg: SafetyConfig α) (s: HRCState α) :=
  minimumPermissibleAction (monitorRisk cfg s)

def runMonitor (cfg: SafetyConfig α) (s: HRCState α): MonitorResult :=
  { risk := monitorRisk cfg s,
    action := monitorAction cfg s }
\end{lstlisting}

The function \texttt{runMonitor} combines the computed risk and the corresponding canonical action into a single monitoring result. Because the monitor is constructed directly from the executable definitions introduced in this section, the same implementation is used in the correctness proofs presented in Section~\ref{sec:correctness}.

Listings~\ref{lst:riskLevel} and~\ref{lst:monitor} therefore expose only the central executable definitions used to explain the monitoring pipeline; the complete implementation and supporting definitions are provided in the accompanying Lean 4 development.



\section{Machine-Checked Correctness}
\label{sec:correctness}

Having constructed the executable runtime monitor, we now establish its
formal correctness. Rather than verifying an abstract specification, the
following theorems are proven directly over the executable Lean
definitions introduced in Section~\ref{sec:monitoring}. 
The objective is not to establish complexity of the individual proofs, but to provide a machine-checked correspondence from executable risk assessment and policy evaluation through action selection to the resulting monitor behaviour.
The verification proceeds incrementally, beginning with policy correctness and extending to behavioural properties and minimal intervention. Collectively, these results demonstrate that the monitor satisfies its intended safety requirements while remaining executable.


\vspace*{-0.2cm}

\subsection{Verification Strategy}
\label{sec:verification-strategy}

The verification follows the layered structure of the monitoring framework. Rather than reasoning about the complete monitor as a monolithic function, we establish properties of the runtime safety policy and use the underlying executable definitions to reason about the behaviour of the monitor and its action selection. This organization separates policy compliance, runtime
behaviour, and intervention minimality into distinct proof obligations. Accordingly, the machine-checked correctness results are organized into three categories:

\begin{itemize}

\item[$\circ$] \textbf{Policy correctness} establishes that every action selected by the monitor satisfies the verified runtime safety policy.

\item[$\circ$] \textbf{Behavioural correctness} characterizes the observable behaviour of the monitor, including deterministic action selection, risk-responsive behaviour, and prevention of unsafe continuation.

\item[$\circ$] \textbf{Minimal intervention} proves that the monitor always selects the least restrictive action permitted by the runtime safety policy.

\end{itemize}

Together, these properties establish that the executable monitor is policy compliant, responds consistently to the assessed runtime risk, and avoids unnecessarily restrictive interventions. The following subsections
present the principal correctness results together with selected supporting properties. Additional lemmas and the complete machine-checked proofs are available in the accompanying Lean~4 development\footnote{\url{https://github.com/codeforpapers3/leansafehrc}}.


\vspace*{-0.2cm}

\subsection{Policy Correctness}
\label{sec:policy-correctness}

Policy correctness establishes that the executable action-selection procedure never produces an action forbidden by the runtime safety policy. The principal result is stated as follows.

\begin{theorem}[Policy Soundness]
\label{thm:monitor-soundness}
For every safety configuration \(c\) and runtime state \(s\),
\[
\operatorname{allowed}
\bigl(
\operatorname{monitorRisk}(c,s),
\operatorname{monitorAction}(c,s)
\bigr).
\]

\end{theorem}

The theorem follows from the construction of the monitor. The risk-assessment procedure computes the qualitative runtime risk, while the canonical action-selection function always returns an action permitted by the corresponding safety policy. Consequently, every
monitor decision is policy compliant.

The same result can be expressed using the set-theoretic policy representation.

\begin{corollary}[Membership in the Permissible-Action Set]
\label{cor:monitor-membership}
For every safety configuration \(c\) and runtime state \(s\),

\vspace*{-0.1cm}

\[
\operatorname{monitorAction}(c,s)
\in
\operatorname{permissibleActions}
\bigl(
\operatorname{monitorRisk}(c,s)
\bigr).
\]

\end{corollary}

This corollary follows immediately from the equivalence between the logical and set-theoretic policy representations established in Section~\ref{sec:policy}. Together, the theorem and corollary establish that the executable Boolean policy, its logical interpretation, and the set-theoretic representation characterize the same policy-compliant monitor behaviour.


\vspace*{-0.2cm}

\subsection{Behavioural Correctness}
\label{sec:behavioural-correctness}

Behavioural correctness characterizes how the executable monitor responds to the qualitative runtime risks computed by the assessment procedure. The Lean~4 development establishes deterministic action selection together with risk-specific response properties for the different qualitative risk levels. In particular, warning, danger, and emergency classifications
produce \texttt{reduceSpeed}, \texttt{protectiveStop}, and \texttt{emergencyStop}, respectively. Additional supporting lemmas and their machine-checked proofs are available in the accompanying formal development.

A representative consequence of these behavioural properties is that unrestricted robot operation is possible only when the assessed runtime risk is safe.

\begin{theorem}[Safe Continuation]
\label{thm:safe-continuation}
For every safety configuration \(c\) and runtime state \(s\),
\vspace*{-0.2cm}
\[
\operatorname{monitorAction}(c,s)=\operatorname{continueOperation}
\Longrightarrow
\operatorname{monitorRisk}(c,s)=\operatorname{safe}.
\]
Equivalently,
\[
\operatorname{monitorRisk}(c,s)\neq\operatorname{safe}
\Longrightarrow
\operatorname{monitorAction}(c,s)\neq\operatorname{continueOperation}.
\]
\end{theorem}


Since \textit{continueOperation} is associated exclusively with the \texttt{safe} risk level, any warning, danger, or emergency condition necessarily results in a more restrictive mitigation action.


\vspace*{-0.2cm}

\subsection{Minimal Intervention}
\label{sec:minimal-intervention}

Policy correctness guarantees that the monitor always selects an admissible
action, while behavioural correctness characterizes its observable response.
The final correctness property establishes that the selected action is also
optimal with respect to the intervention ordering introduced in
Section~\ref{sec:framework-orderings}. Specifically, the monitor always
chooses the least restrictive action permitted by the runtime safety policy.

\begin{theorem}[Minimal Intervention]
\label{thm:minimal-intervention}

For every safety configuration \(c\), runtime state \(s\), and action
\(a\) satisfying

\vspace*{-0.3cm}

\[
a
\in
\operatorname{permissibleActions}
\bigl(
\operatorname{monitorRisk}(c,s)
\bigr),
\]

\vspace*{-0.2cm}

we have

\vspace*{-0.4cm}

\[
\operatorname{actionSeverity}
\bigl(
\operatorname{monitorAction}(c,s)
\bigr)
\le
\operatorname{actionSeverity}(a).
\]

\vspace*{-0.4cm}

\end{theorem}

The result follows from the canonical action-selection procedure. The selected monitor action is the least restrictive action permitted by the verified policy; hence, no admissible action with lower intervention
severity can exist. Together with Policy Soundness
(Theorem~\ref{thm:monitor-soundness}), this establishes that the selected action is both policy compliant and minimally restrictive relative to the formalized policy and action ordering.

Collectively, the machine-checked results establish policy compliance, deterministic and risk-responsive behaviour, prevention of unsafe continuation, and minimal intervention for the executable monitor. The
following section demonstrates these properties in a collaborative assembly case study.


\vspace*{-0.4cm}

\section{Collaborative Assembly Case Study}
\label{sec:case-study}

\vspace*{-0.2cm}

We demonstrate the executable behaviour of \framework\ through a representative collaborative assembly case study. The case study exercises the monitor across representative operating conditions ranging from normal
collaboration to emergency situations and relates the resulting monitoring decisions to the machine-checked correctness guarantees established in Section~\ref{sec:correctness}.


\vspace*{-0.4cm}

\subsection{Case-Study Setup}
\label{sec:experimental-setup}

The case study uses a fixed safety configuration and seven representative runtime states that collectively exercise all qualitative risk levels and their corresponding canonical mitigation actions. The scenarios cover robot-only operation, safe human--robot collaboration, proximity-based warning and danger conditions, velocity and interaction-force violations,
and an emergency condition. For each runtime state, the executable monitor computes the qualitative risk, selects the corresponding canonical action,
and returns the complete monitoring result. These results are then examined with respect to the machine-checked correctness properties established in
Section~\ref{sec:correctness}. The complete scenario definitions, executable evaluations, and associated proofs are available in the Lean~4 development.


\vspace*{-0.4cm}

\subsection{Evaluation Results}
\label{sec:evaluation}

\vspace*{-0.2cm}

The executable monitor was evaluated over the seven runtime states defined in Section~\ref{sec:experimental-setup}. Table~\ref{tab:case-study-results} summarizes the resulting risk classifications and canonical actions. The complete risk and action sequences are also established by machine-checked Lean~4 theorems, linking executable evaluation directly to formal verification.


\begin{table*}[!ht]
\centering
\caption{Execution of the Verified Runtime Monitor for Representative Collaborative Assembly Scenarios}
\label{tab:case-study-results}
\scriptsize
\renewcommand{\arraystretch}{1.0}
\setlength{\tabcolsep}{4pt}
\begin{tabular}{p{1.0cm}p{0.9cm}p{1.0cm}p{0.9cm}p{0.8cm}p{1.3cm}p{1.1cm}p{1.2cm}p{1.6cm}}
\toprule
\textbf{Scenario}
&
\textbf{Human}
&
\textbf{Distance}
&
\textbf{Velocity}
&
\textbf{Force}
&
\textbf{Emergency}
&
\textbf{Risk}
&
\textbf{Monitor Action}
&
\textbf{Scenario Characteristic}
\\
\midrule

S1 &
Absent &
N/A &
Normal &
Normal &
No &
Safe &
Continue Operation &
Human \ \ \ \ \ Absent
\\

S2 &
Present &
Safe &
Normal &
Normal &
No &
Safe &
Continue Operation &
Safe Collaboration
\\

S3 &
Present &
Warning &
Normal &
Normal &
No &
Warning &
Reduce Speed &
Proximity Warning
\\

S4 &
Present &
Safe &
High &
Normal &
No &
Warning &
Reduce Speed &
Velocity \ \ \ \ \ Violation
\\

S5 &
Present &
Danger &
Normal &
Normal &
No &
Danger &
Protective Stop &
Proximity Danger
\\

S6 &
Present &
Safe &
Normal &
High &
No &
Danger &
Protective Stop &
Force \ \ \ \ \ \ \ \ \ \ Violation
\\

S7 &
Present &
Safe &
Normal &
Normal &
Yes &
Emergency &
Emergency Stop &
Emergency Override
\\

\bottomrule
\end{tabular}
\end{table*}


The execution results exhibit the expected progression of monitor responses. Scenarios~S1--S2 permit continued operation under safe conditions; S3--S4 result in speed reduction, S5--S6 result in protective stop, and S7 results in emergency stop. The priority-based assessment is illustrated particularly by S4 and S6: a velocity violation upgrades an otherwise safe geometric assessment
to \texttt{warning}, while an excessive interaction force upgrades an otherwise safe geometric assessment to \texttt{danger}. Scenario~S7 demonstrates emergency dominance. Across the complete trace, the computed risk sequence is \texttt{safe}, \texttt{safe}, \texttt{warning}, \texttt{warning}, \texttt{danger}, \texttt{danger}, and \texttt{emergency}, with the
corresponding canonical actions established by machine-checked Lean~4 theorems.


\vspace*{-0.7cm}

\subsection{Case Study Analysis}
\label{sec:case-study-analysis}

\vspace*{-0.4cm}

The case study provides both executable instances and machine-checked
instantiations of the general correctness results from
Section~\ref{sec:correctness}. In addition to establishing the complete
risk and action sequences, the Lean~4 development establishes policy
soundness for representative safe, warning, danger, and emergency states.
It further proves that representative danger, excessive-force, and
emergency scenarios cannot result in unrestricted continuation, and that
an emergency classification necessarily produces \texttt{emergencyStop}.

The case-study development also instantiates the minimal-intervention
theorem for warning, danger, and emergency conditions. Consequently,
\texttt{reduceSpeed}, \texttt{protectiveStop}, and
\texttt{emergencyStop} are respectively established as the least
restrictive permissible responses for the corresponding representative
states. These results show that the case study is not merely an execution
trace: its monitoring decisions are directly connected to the generic
correctness theorems of the framework through machine-checked proofs.

Although demonstrated using collaborative assembly, the monitoring
architecture is not tied to this particular scenario. Other HRC
applications with compatible risk and action models can instantiate
application-specific runtime states and safety configurations while
reusing the executable monitoring pipeline and its generic correctness
results.


\vspace*{-0.4cm}

\section{Discussion}
\label{sec:discussion}

\vspace*{-0.2cm}

The proposed framework demonstrates that executable runtime monitoring and
machine-checked verification can be developed within a single Lean~4
formalization. Rather than treating implementation and verification as
separate activities, \framework\ represents runtime risk assessment, policy
evaluation, action selection, and monitor construction as executable
definitions that also serve as the basis for machine-checked reasoning.
This avoids a separate monitor implementation whose correspondence with the
verified model would itself require justification, while providing
correctness guarantees beyond those obtainable from testing or simulation
alone.

Beyond the collaborative assembly case study, the framework provides a
reusable monitoring architecture for HRC applications with compatible risk
and action models. Such applications can instantiate application-specific
runtime states and safety configurations while reusing the executable
monitoring pipeline and its generic correctness results. Applications
requiring different risk structures, mitigation policies, or decision
semantics would require corresponding extensions to the formal model and
their associated proofs.

The current work intentionally focuses on qualitative runtime risk
assessment using configurable safety thresholds, with runtime states
evaluated independently through executable Lean~4 scenarios. The monitor
has not yet been integrated with a physical robotic platform or live sensor
stream. Future work will investigate quantitative and probabilistic risk
models, temporal and predictive monitoring, and physical deployment through
robotics middleware such as ROS~2. Such extensions would require
corresponding machine-checked correctness arguments for the additional
monitoring functionality.


\vspace*{-0.5cm}

\section{Conclusion}
\label{sec:conclusion}

\vspace*{-0.3cm}

This paper presented \framework, a Lean~4 framework for verified runtime
safety monitoring in human--robot collaboration. The framework integrates
runtime risk assessment, policy evaluation, canonical action selection,
and monitor construction within a unified formal development. The same
executable definitions used to construct the monitor also serve as the
basis for its machine-checked correctness proofs, avoiding a separate
implementation layer whose correspondence with the verified model would
require additional justification.
The framework establishes machine-checked guarantees of policy soundness,
deterministic and risk-responsive behaviour, prevention of unsafe
continuation, and minimal intervention. A representative collaborative
assembly case study demonstrates the executable realization of these
properties and provides machine-checked instantiations of the general
correctness results. 
theorem proving can support the construction of executable runtime monitors
with rigorous correctness guarantees for safety-critical robotic systems.

Future work will extend the framework with quantitative and
uncertainty-aware risk assessment, temporal and predictive monitoring,
integration with robotic middleware such as ROS~2, and deployment on
physical collaborative robotic platforms. We also plan to investigate the
application of the proposed methodology to other systems requiring trustworthy runtime monitoring.

\bibliographystyle{splncs04}
\bibliography{references}


\end{document}